\documentclass[11pt, a4paper]{article}
\usepackage{moriond,epsfig}
\def\be{\begin{equation}}
\def\ee{\end{equation}}
\def\bea{\begin{eqnarray}}
\def\eea{\end{eqnarray}}

\def\mWeq{M_{W}}
\def\mW{$\mWeq$}

\def\mt{$M_{t}$}
\def\mH{$M_{H}$}
\def\gW{$\Gamma_{W}$}
\def\enu{$e\nu$}
\def\munu{$\mu\nu$}
\def\mTeq{M_{T}}
\def\mT{$\mTeq$}
\def\pt{$p_{T}$}
\def\ptleq{p_{T}^{\ell}}
\def\ptl{$\ptleq$}
\def\ptlv{$\vec{\ptleq}$}
\def\u{$\vec{U}$}

\def\Wenu{$W\rightarrow e\nu$}
\def\Wmunu{$W\rightarrow\mu\nu$}
\def\Wlnu{$W\rightarrow\ell\nu$}
\def\Wtaunu{$W\rightarrow\tau\nu$}
\def\Zll{$Z\rightarrow\ell\ell$}
\def\Ztautau{$Z\rightarrow\tau\tau$}
\def\metEq{{E}\!\!\!/_T }
\def\met{$\metEq$}
\def\ipb{$\rm pb^{-1}$}
\begin{document}
\vspace*{4cm}
\title{W mass and width measurements at the Tevatron}

% Changed 040623 - All authors go here
\author{Emily Nurse (for the CDF collaboration)}
% 
%\address{(1) Laboratoire de l'Accelerateur Lineaire\\
%Universite Paris Sud 11, BP 34, 91898 Orsay Cedex, France}
% 
\address{University College London, Gower Street, London, WC1E 6BT, United Kingdom.}

\maketitle\abstracts{ 
I present a measurement of the W boson mass (\mW) and width (\gW) using 200 and 350~\ipb\ of CDF Run II data respectively. 
The measurements, performed in both the electron and muon decay channels, rely on a fit to the W transverse mass distribution.
We measure \mW\ = 80413 $\pm$ 48~MeV and  \gW\ = 2032 $\pm$ 71~MeV which represent the world's single most precise measurements to date.}
\section{Introduction}
%\be
%\mWeq^{2} = \frac{\pi \alpha \mZeq^{2}}{\sqrt{2}G_{F}(1 - (\frac{\mZeq}{\mWeq}))} \frac{1}{(1-\Delta r)}
%\label{eq:mw}
%\ee
The mass (\mW) and width (\gW) of the W boson are important parameters of the Standard Model (SM).
Radiative corrections to the W propogator are dominated by Higgs and top-bottom loops, thus a precise
measurement of \mW\ together with \mt, the mass of the top quark, place an indirect constraint on the 
mass of the as yet un-discovered Higgs boson, \mH. 
A precise measurement of \gW\ provides a stringent test of the SM prediction which is accurate to 2~MeV~\cite{pdg}.

At the Tevatron W bosons are predominantly produced via quark anti-quark annihilation. 
The measurements are performed in the \enu\ and \munu\ decay channels which provide clean experimental 
signatures. The \mW\ and \gW\ analyses utilise 200~\ipb\ and 350~\ipb\ of CDF data from Run II at the Tevatron 
respectively. 
%The analyses were performed independently and although the analysis techniques are very
%similar the dominant systematics are different. Minimisation of the total uncertainty thus requires some
%things to be 

Since neutrinos are not detected in CDF the W invariant mass cannot be reconstructed. 
Instead we reconstruct the transverse mass, \mT, which is defined as:
% in Eq.~\ref{eq:mt}.
\be
\mTeq = \sqrt{2\ptleq p_{T}^{\nu}(1 - \cos \phi_{\ell\nu})}
\label{eq:mt}
\ee
where \ptl\ is the transverse momentum (\pt) of the charged lepton, $p_{T}^{\nu}$ is
the \pt\ of the neutrino and $\phi_{\ell\nu}$ is the azimuthal angle between the charged lepton and the
neutrino. $p_{T}^{\nu}$ is inferred from the transverse momentum imbalance in the event.

A Monte Carlo simulation is used to predict the \mT\ distribution as a function of \mW\ and \gW.
These predictions are fitted to the data with a binned maximum-likelihood fit in order to extract
 \mW\ and \gW. The fit for \mW\ is performed in the region around the peak of the distribution: 65--90~GeV. The fit
for \gW\ is performed in the high \mT\ tail region: 90--200~GeV, which is still sensitive to the Breit-Wigner
line-shape but less sensitive to the Gaussian detector resolutions.
These line-shape predictions depend on a number of production and detector effects.
The most important effects are described in this document 
and all the systematic uncertainties are summarised at the end.
\section{Monte Carlo Simulation}
A dedicated parameterised Monte Carlo simulation is used to generate the  \mT\ templates used in the fits.
The W \pt\ spectrum is modelled with RESBOS~\cite{resbos} and QED corrections for one photon emission are simulated with 
Berends and Kleiss~\cite{bk} and WGRAD.~\cite{wgrad}  Systematic uncertainties arise from non-perturbative QCD parameters
affecting the W \pt\ spectrum and considerations of the emission of a second photon from the final state charged lepton.
Parton distribution functions (PDFs) affect the acceptance and kinematics of decay products. The templates are
generated with the CTEQ6M~\cite{cteq} PDFs and their error sets are used to estimate the PDF uncertainty.

The detector response model is tuned to \Zll\ and \Wlnu\ data as well as a full GEANT Monte Carlo
simulation of the CDF detector.
%\section{Lepton Momentum calibration: Scale and Resolution}
\section{Lepton Calibration: Scales and Resolutions}
The muon momentum is measured in a cylindrical drift chamber. The scale and resolution of the momentum
are calibrated using the resonance peaks in $J/\Psi\rightarrow\mu\mu$, $\Upsilon(1S)\rightarrow\mu\mu$ and $Z \rightarrow\mu\mu$
events utilising the precisely measured world average masses of these particles.~\cite{pdg} The $J/\Psi$ sample has sufficient
statistics to verify the linearity of the momentum scale by studying its variance as a function of muon \pt.
Combining all three measurements enables an accuracy of 0.021\%\ on the momentum scale. 
%\section{Electron Energy calibration: Scale and Resolution}

The electron energy is measured in the calorimeter. The electron momentum is also measured in the drift
chamber~\footnote{Since the mass of the electron is negligible the true momentum and energy values are the same. 
                  However collinear photon radiation from the electron which is clustered back into the energy 
                  measurement can decrease the track momentum measurement. These effects are well modelled in the simulation.},
thus the well calibrated momentum measurement is used to calibrate the calorimeter scale (response) and resolution
using the ratio between the electron energy measured in the calorimeter and the track momentum (E/p)
in \Wenu\ events. The scale and resolution can also be obtained independently from the mass peak in $Z \rightarrow ee$ events.
The two measurements are combined to give a calorimeter scale measurement accurate to 0.034\%.
\section{Hadronic Recoil Calibration}
The neutrino \pt\ is determined from the missing transverse energy, \met, in the detector. 
A recoil vector, \u, is defined as the vector sum of transverse energy  over all calorimeter towers, excluding those surrounding the lepton. 
The \met\ is then defined as -(\u\ + \ptlv).
The recoil has contributions from initial state gluon radiation from the incoming quarks, underlying event energy and final state photon
radiation from the charged lepton. The recoil is represented by a parameterised model, which is tuned in \Zll\ events.
%The recoil model splits \u\ into two components, \uone\ and \utwo\
%which are parallel and perpendicular to the Z \pt\ direction respectively. Most of the recoil is in the direction anti-parallel to the Z \pt\ since
%the Z is recoiling against the gluons from initial state radiation. 
The model parameters are found from the Z data and applied to the W data. The systematic uncertainties on \mW\ and \gW\ come from 
the uncertainties on the model parameters due to the limited statistics in the Z data.
%In W events the W \pt\ cannot be reconstructed due to the undetected neutrino
\section{Backgrounds}
Backgrounds have different \mT\ distributions to \Wlnu\ events, therefore the \mT\ shape must be added to the Monte Carlo templates when
fitting to the data. Electroweak backgrounds consist of \Zll\ events where one of the leptons goes undetected and \Wtaunu\ and \Ztautau\ 
events where the $\tau$ decays to an electron or muon. 
These backgrounds are found using Pythia~\cite{pythia} Monte Carlo samples of W and Z events, passed through
a full GEANT simulation of the CDF detector. Non-electroweak backgrounds consist of multi-jet events, where one jet fakes or contains a 
lepton and the other is sufficiently mis-measured to produce \met, and (in the muon channel only) kaons that decay to muons within 
the volume of the drift chamber. In the latter case the resulting reconstructed track contains a kink that can produce a fake high measured \pt\ and \met.
The multi-jet background normalisations are found by fitting the low \met\ distribution where this background dominates. 
The \mT\ distributions are found by reversing certain lepton identification cuts.
The kaon background is found by fitting the high tail of the track fit $\chi^2$ distribution where this background is large. The 
\mT\ shape is found by reversing an impact parameter cut. 
%Figure xx shows the \mt\ shapes and normalisiations.
\section{Results}
\begin{figure}
%\rule{5cm}{0.2mm}\hfill\rule{5cm}{0.2mm}
%\vskip 2.5cm
%\rule{5cm}{0.2mm}\hfill\rule{5cm}{0.2mm}
\epsfig{figure=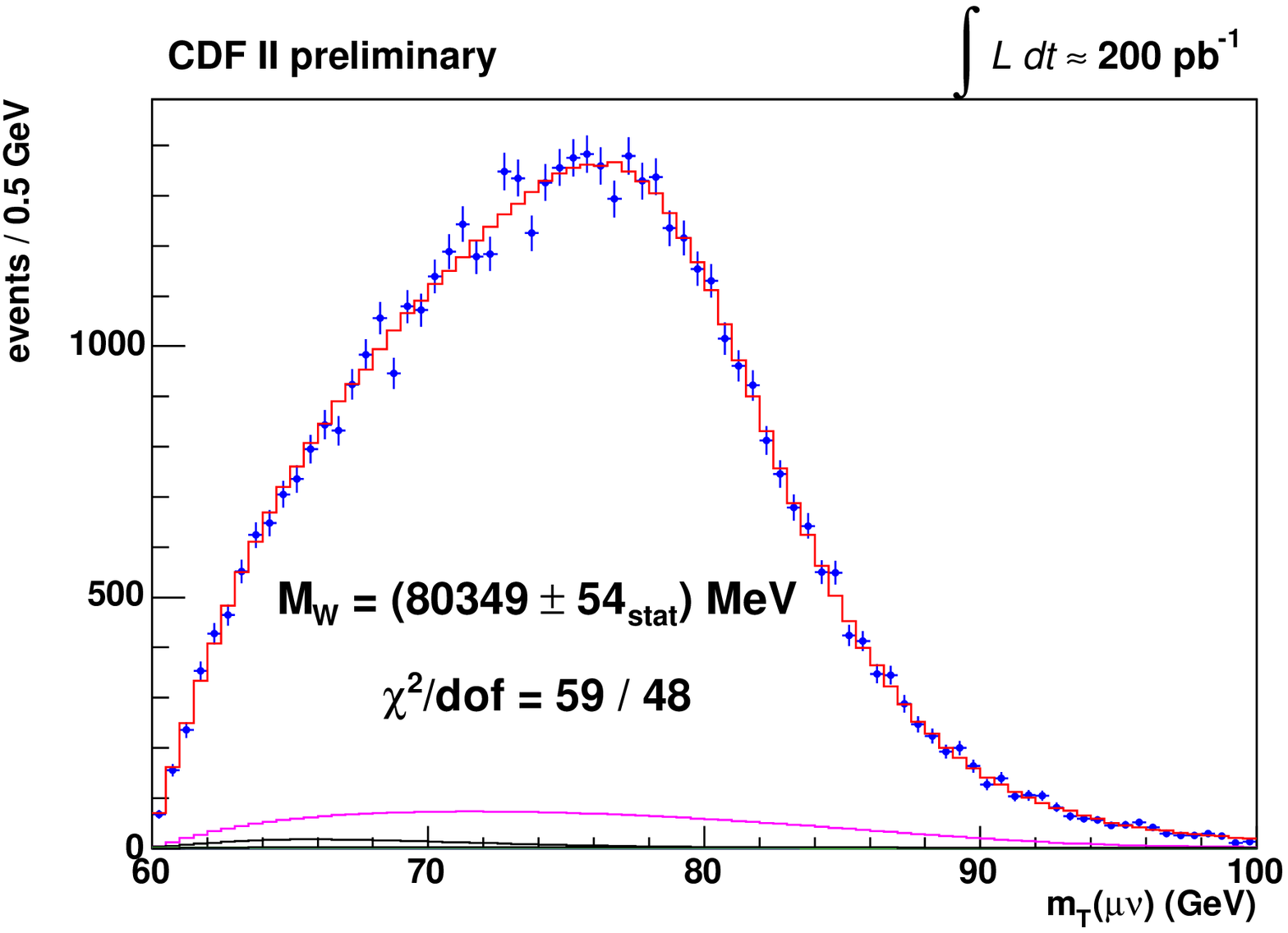,height=6.5cm,width=8cm}
\epsfig{figure=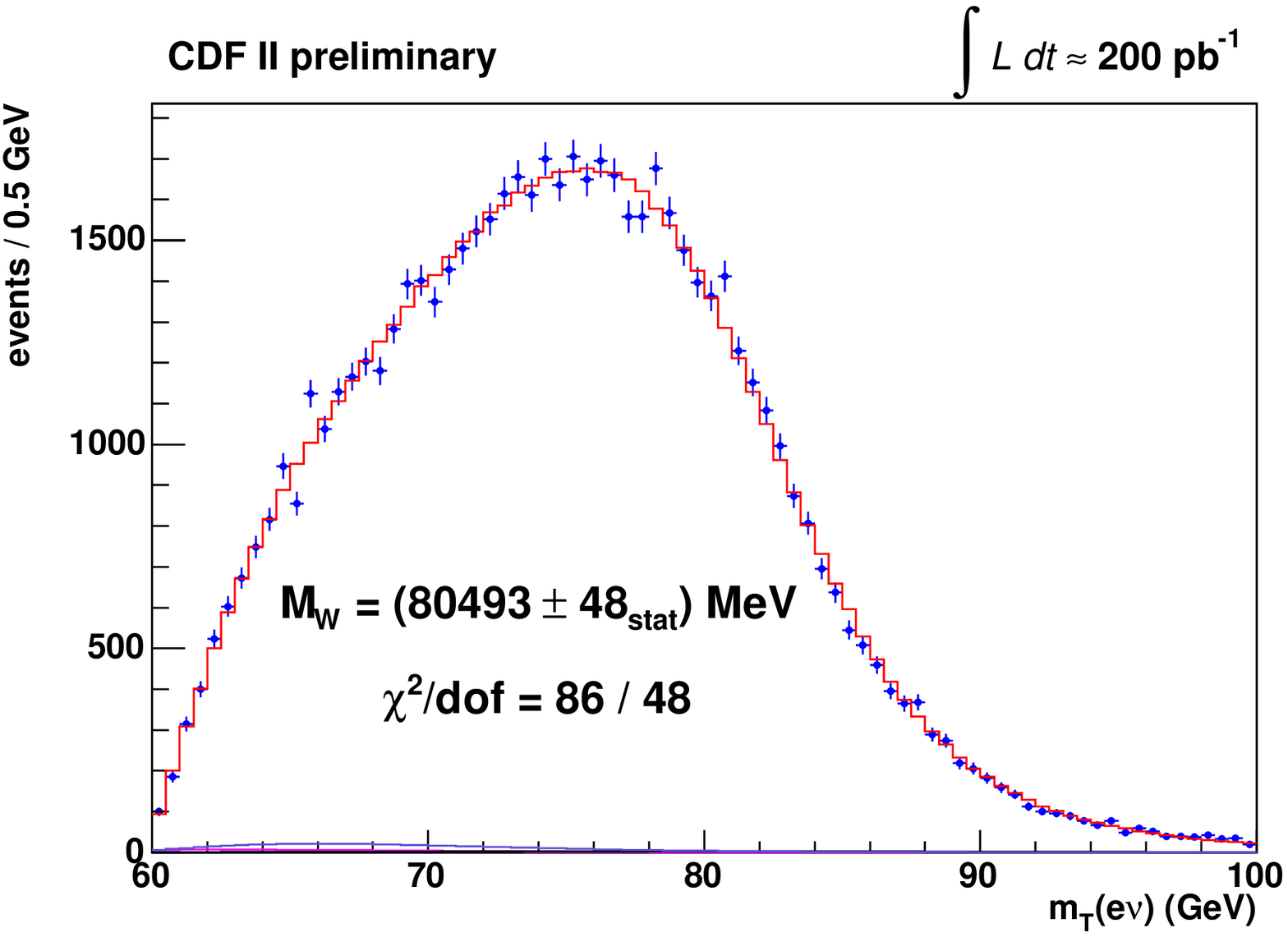,height=6.5cm,width=8cm}
\caption{Transverse mass fits for \mW\ in \Wmunu\ (left) and \Wenu\ (right) events. The fit is performed in the region 65--90~GeV.
\label{fig:mass-fits}}
\end{figure}

\begin{figure}
%\rule{5cm}{0.2mm}\hfill\rule{5cm}{0.2mm}
%\vskip 2.5cm
%\rule{5cm}{0.2mm}\hfill\rule{5cm}{0.2mm}
\epsfig{figure=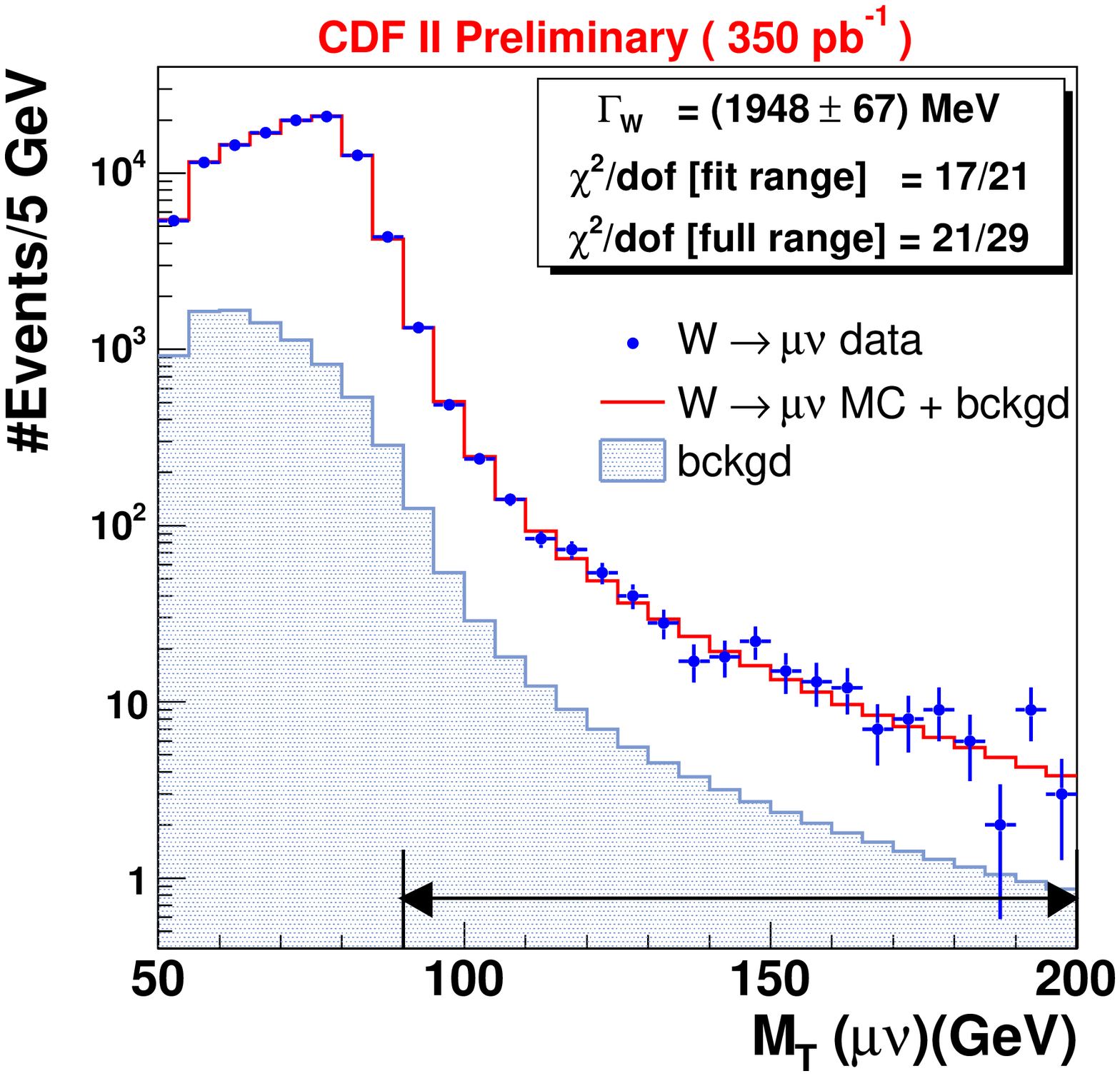,height=7cm,width=8cm}
\epsfig{figure=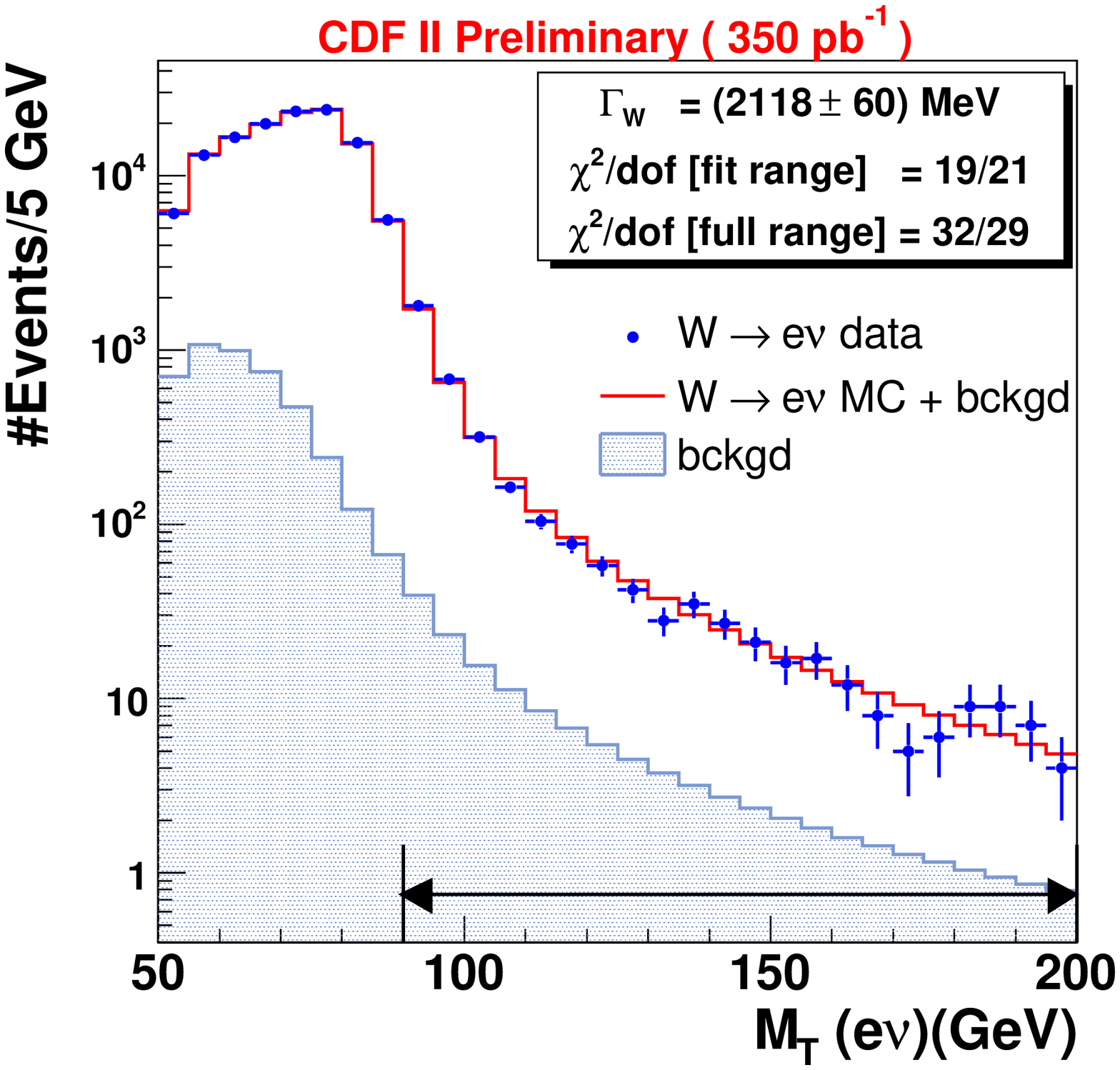,height=7cm,width=8cm}
\caption{Transverse mass fits for \gW\ in \Wmunu\ (left) and \Wenu\ (right) events. The fit is performed in the region 90--200~GeV.
\label{fig:width-fits}}
\end{figure}

The systematic and statistical uncertainties for \mW\ and \gW\ are summarised in Table~\ref{tab:mass}.
Figure~\ref{fig:mass-fits} shows the \mT\ fits for \mW\ in the muon and electron decay channels. The fitted \mW\ values are combined together with fits to
the charged lepton \pt\ and \met\ distributions to give \mW\ = 80413 $\pm$ 48~MeV, the world's most precise single measurement. This result increases the
world average central value by 6~MeV and reduces the uncertainty by 15\%. The updated world average impacts the global precision electroweak fits, reducing
the preferred \mH\ by 6~GeV to $76^{+33}_{-24}$~GeV. 
The 95\%\ CL upper limit on \mH\ is 144(182)~GeV with(out) the LEP~II direct limit included.~\cite{lep,renton-grunewald}
Figure~\ref{fig:width-fits} shows the  \mT\ fits for \gW\ in the  muon and electron decay channels. The results are combined to give the final result 
\gW\ = 2032 $\pm$ 71~MeV, the world's most precise single measurement, which is in good agreement with the SM prediction. This result reduces the
world average central value by 44~MeV and uncertainty by 22\%.
%Figure~\ref{fig:results} shows the new \mW\ and \gW\ measurements compared to current experimental measurements and (for \gW) the SM prediction.
\begin{table}[t]
\caption{Uncertainties for the W mass (left) and W width (right). The third column lists the uncertainties that are common between the electron and muon channels.\label{tab:mass}}
\vspace{0.4cm}
\begin{center}
\begin{tabular}{|l|c|c|c|}
\hline
%& & & \\
$\Delta$\mW\ [MeV]      & $e$       & $\mu$  &  C     \\ \hline
Lepton Scale            &  30       & 17     & 17     \\
Lepton Resolution       &  ~9       & ~3     & ~0     \\
Recoil Scale            &  ~9       & ~9     & ~9     \\
Recoil Resolution       &  ~7       & ~7     & ~7     \\
Lepton ID               &  ~3       & ~1     & ~0     \\
Lepton Removal          &  ~8       & ~5     & ~5     \\
Backgrounds             &  ~8       & ~9     & ~0     \\
\pt(W)                  &  ~3       & ~3     & ~3     \\
PDF                     &  11       & 11     & 11     \\
QED                     &  11       & 12     & 11     \\ \hline
Total Systematic        &  39       & 27     & 26     \\ \hline
Statistical             &  48       & 54     & ~0     \\ \hline \hline
Total                   &  62       & 60     & 26     \\ \hline
\end{tabular}
\hspace{1cm}
\begin{tabular}{|l|c|c|c|}
\hline
%& & & \\
$\Delta$ \gW\ [MeV]     & $e$      &  $\mu$ &   C     \\ \hline
Lepton Scale            &  21       & 17     & 12     \\
Lepton Resolution       &  31       & 26     & ~0     \\
Simulation              &  13       & ~0     & ~0     \\
Recoil                  &  54       & 49     & ~0     \\
Lepton ID               &  10       & ~7     & ~0     \\
Backgrounds             &  32       & 33     & ~0     \\
\pt(W)                  &  ~7       & ~7     & ~7     \\
PDF                     &  16       & 17     & 16     \\
QED                     &  ~8       & ~1     & ~1     \\ 
\mW\                    &  ~9       & ~9     & ~9     \\ \hline  
Total Systematic        &  78       & 70     & 23     \\ \hline
Statistical             &  60       & 67     & ~0     \\ \hline \hline
Total                   &  98       & 97     & 23     \\ \hline
\end{tabular}
\end{center}
\end{table}

%\clearpage

%\begin{figure}
%\vskip 2.5cm
%\epsfig{figure=mwsummary.eps,height=6.5cm,width=8cm}
%\epsfig{figure=GammaW_Comparison_42-2.eps,height=6.5cm,width=8cm}
%\caption{The new \mW\ (left) and \gW\ (right) measurements compared to the current experimental measurements and (for \gW) the SM prediction.}
%\label{fig:results}
%\end{figure}

\end{document}